\documentclass[12pt]{article}
\newcommand{\doublespacing}{\baselineskip=26pt plus 0.5pt minus 1pt}
\usepackage[dvips]{graphicx}
\font\scaps=cmcsc10    % small capitals
\font\cal=cmsy10    % calligraphic
\newcommand \be{\begin{equation}}
\newcommand \ee{\end{equation}}
\newcommand \ba{\begin{eqnarray}}
\newcommand \ea{\end{eqnarray}}

\begin{document}

\setkeys{Gin}{draft=false}    % to produce draft
\font\bgr=cmmib10 % bold greek
\font\bfgreek=cmmib10 \textfont9=\bfgreek

\def\today{\ifcase\month\or
  January\or February\or March\or April\or May\or June\or
  July\or August\or September\or October\or November\or December\fi
  \space\number\day, \number\year}
%
%\hfil PostScript file created: \today{}; \ time \the\time \ minutes
\vskip .05in

\title{On geometric complexity of earthquake focal
zone and fault system: A statistical study }

\author{Yan Y. Kagan$^1$\footnote{Yan Y. Kagan, Department of
Earth and Space Sciences, University of California, Los
Angeles, California, 90095-1567, USA; (e-mail:
ykagan@ucla.edu)}
\\ $^1$ Department of Earth and Space Sciences\\ University of
California, Los Angeles, California, USA} \maketitle
\begin{abstract}
\doublespacing We discuss various methods used to investigate the
geometric complexity of earthquakes and earthquake faults, based
both on a point-source representation and the study of
interrelations between earthquake focal mechanisms. We briefly
review the seismic moment tensor formalism and discuss in some
detail the representation of double-couple (DC) earthquake sources
by normalized quaternions. Non-DC earthquake sources like the CLVD
focal mechanism are also considered. We obtain the characterization
of the earthquake complex source caused by summation of disoriented
DC sources. We show that commonly defined geometrical fault barriers
correspond to the sources without any CLVD component. We analyze the
CMT global earthquake catalog to examine whether the focal mechanism
distribution suggests that the CLVD component is likely to be zero
in tectonic earthquakes. Although some indications support this
conjecture, we need more extensive and significantly more accurate
data to answer this question fully.

\vskip 0.1 truein

{\scaps {Keywords}: } Earthquake focal mechanism, double couple,
CLVD, quaternion, geometric barriers, statistical analysis

\end{abstract}

%\begin{article}
\doublespacing
% \clearpage
\newpage

\section{Introduction}
\label{intr}

Several statistical methods can be used to study the geometric
complexity of an earthquake fault zone or fault system. Some are
based on representing an earthquake as a point source, and the
geometric complexity of a source reveals itself in a complex
structure of a seismic moment tensor. Another method is to
investigate the geometric complexity of the fault system as
expressed in a set of earthquake locations and their focal
mechanisms.

\subsection {Models for complex earthquake sources }
\label{intr1} \hfil\break
\subsubsection {Phenomenological observations }
\label{intr11} \hfil\break $\bullet$ 1. Geologic and geophysical
studies of earthquake focal zones point to significant complexity in
the rupture process (see, for example, King, 1983, 1986). Moreover,
almost any large earthquake is now analyzed in detail with its
rupture history represented in time-space-focal mechanism maps which
usually exhibit an intricate moment release. Although such results
suggest that an earthquake focal zone is more complex than the
standard planar model (Aki and Richards, 2002), phenomenological
investigations cannot describe the fault patterns appropriate for
its quantitative modeling. Such a description needs to be based on
statistical treatment of the observed fault geometry.

\hfil\break
\subsubsection {Point source solutions }
\label{intr12} \hfil\break $\bullet$ 2. As a first approximation,
the moment tensor is a second-rank matrix (Backus \& Mulcahy,
1976a,b; Backus, 1977a,b; Kagan, 1987). The main component of the
second-rank tensor is a double-couple (Burridge and Knopoff, 1964).
The presence of a significant non-double-couple or a CLVD component
(Knopoff and Randall, 1970) is one measure of source complexity
(Section~\ref{YYK_gam}). In this work we assume that for earthquakes
the isotropic component equals zero.

Julian {\it et al.}\ (1998) and Miller {\it et al.}\ (1998) discuss
various physical mechanisms thought to be responsible for the non-DC
and, in particular, CLVD earthquake sources. They review many
published papers that describe the registration of the CLVD
component. Based on theoretical considerations, they indicate that
most of the CLVD cases should be seen in geothermal and volcanic
areas and quote observations confirming this. However, the
complexity of an earthquake source can also produce a non-zero CLVD
component in some tectonic earthquakes. Thus, the CLVD evaluation by
modern moment tensor solutions should, in principle, quantitatively
characterize the complexity of the earthquake fault zone. Julian
{\it et al.}\ (1998) and Miller {\it et al.}\ (1998) provide many
published examples of a significant CLVD component present in
tectonic earthquakes.

However, recent work by Frohlich and Davis (1999), Kagan (2003), and
Frohlich (2006, pp.~228-235) appears sceptical that present
inversion techniques can obtain an accurate estimate of the CLVD
component for tectonic earthquakes. Fig.~16 in Miller {\it et al.}\
(1998) also demonstrates the lack of correlation between the CLVD
component values obtained from different earthquake catalogs,
indicating that these non-zero CLVD values may come from systematic
effects. These results suggest that routinely determined CLVD-values
would not reliably show the deviation of earthquake focal mechanisms
from a standard DC model. \hfil\break $\bullet$ 3. Higher-rank point
seismic moment tensors were introduced by Backus and Mulcahy
(1976a,b) and Backus (1977a,b). Silver and Jordan (1983), Silver and
Masuda (1985), Kagan (1987, 1988), McGuire {\it et al.}\ (2001),
Chen {\it et al.}\ (2005) considered various aspects of higher-rank
point seismic moment tensors. Kagan (1987) argued that the
third-rank seismic moment tensor should show the complexity of the
earthquake source, i.e., its difference from the standard planar
rupture model. However, with the data available now, inversion
results indicate that only the extent and directivity of the rupture
can be obtained by analyzing higher-rank tensors.

\hfil\break
\subsubsection {Set of earthquake focal mechanisms}
\label{intr13} \hfil\break $\bullet$ 4. Study of higher-rank
correlation tensors (Kagan and Knopoff, 1985b; Kagan, 1992a). These
correlation tensors more completely describe the interrelation
between focal mechanisms and their spatial distribution. However,
their interpretation is still difficult and the low accuracy of
available catalogs makes conclusions uncertain. Below we consider
one of the invariants of the correlation tensor (tensor dot product
for two arbitrary solutions). \hfil\break $\bullet$ 5. Investigation
of the geometric relations between the double-couple earthquake
mechanism solutions (Kagan, 1982, 1990, 1991, 1992b, 2000, 2005).
These papers studied the interrelation between the pairs of
earthquake focal mechanisms by determining the 3-D angle between two
DC solutions (Section~\ref{YYK_quat3.2.2}) and considered the angle
distribution in space-time. It was shown that the angle is
distributed according to the Cauchy law and increases with time and
distance interval between earthquakes. However, as we will see
later, the fault complexity depends not only on the pairwise
distribution of 3-D rotation angles, but also on the distribution of
rotation poles on a reference unit sphere.

\subsection {Geometric complexity of earthquake faulting }
\label{intr2} \hfil\break There is similarity between the geometric
complexity of the earthquake fault zone and that of the earthquake
fault system. Their comparability is the result of a general
self-similarity of earthquake occurrence: earthquake rupture
propagates over a complex fault pattern. This pattern is then seen
in occurrence of aftershocks and other dependent events. Therefore,
we may assume that the fault pattern complexity, when considered for
small time intervals, would approach the geometric complexity of
each earthquake rupture. Fractal features of the spatial
distribution of earthquake hypocenters and epicenters (Kagan, 2007a)
support this conjecture.

An important condition which contributes to the complexity of the
earthquake fault system is the compatibility of elastic
displacement: there should be no voids or material overlap created
in fault configurations (Gabrielov {\it et al.}, 1996). The authors
developed a mathematical framework for calculating the kinematic and
geometric incompatibility in tectonic block systems, both rigid and
deformable. They concluded that due to geometric incompatibilities
at fault junctions, new ruptures must be created to accommodate
large plate tectonic deformations.

Historically, earthquake fault mechanics was patterned along the
lines of engineering fracture mechanics (see, for example, Anderson,
2005), where the tensile fracture (formation of cracks) in materials
and bodies is a major concern. Voids are created in tensile cracks;
therefore the displacement incompatibility condition is not
satisfied. However, earthquakes occur in Earth's interior, where a
considerable lithostatic pressure should prevent the appearance of
voids.

Moreover, engineering concepts as opposed to physical theories
generally cannot be transferred into a new scientific field without
major experimental work. This may explain why fracture mechanics did
not significantly improve our understanding of earthquake rupture
process (Kagan, 2006).

Similarly, another engineering science discipline, the study of
friction, is based on experiments with man-made surfaces. Though
widely used in earthquake mechanics, its contribution to the
theoretical foundations of this science is still uncertain ({\it
ibid}).

Here we concentrate on analyzing the CLVD component of the seismic
moment tensor. The CLVD is the simplest measure of earthquake focal
zone complexity. Only in Section~\ref{YYK_mdot} do we briefly review
the dot product for seismic moment tensors. This product is the
first invariant of the second-rank correlation tensor (see
Section~\ref{intr13}) and can also be used to characterize the
complexity of the focal mechanism orientation. Further detailed
study needs to address the other two invariants of the correlation
tensor.

Many measurements of focal zone geometry indicate that a planar
earthquake fault is only a first approximation; rupture is usually
non-planar. However, it is important to know whether the focal zone
of a single earthquake or the fault systems of many earthquakes can
be represented by a distribution of small dislocations with the DC
mechanism. If no CLVD component is present in tectonic earthquakes,
one degree of freedom for each rupture patch can be excluded with a
great savings in representing earthquake rupture patterns.

Frohlich {\it et al.}\ (1989) and Frohlich (1990) studied the CLVD
distribution for earthquakes with the rupture slip along surfaces of
revolution and found that a certain geometric slip pattern produces
a significant CLVD component. However, such smooth surfaces of
revolution are unlikely during a real earthquake rupture. Our
results (Kagan, 1992, 2000, 2007a) rather suggest that both the
fault rupture system and focal mechanisms, associated with the
rupture, are non-smooth everywhere. They are controlled by fractal
distributions.

Below, in Section~\ref{YYK_inv} and Section~\ref{YYK_quat} we review
the seismic moment tensor formalism and quaternion representation of
the DC focal mechanism. Section~\ref{YYK_gam} discusses the CLVD
component of a focal mechanism and theoretically evaluates the
component for some models of the earthquake composite source. We use
the global CMT catalog to investigate statistically
(Section~\ref{YYK_stat}) the spatial pattern of earthquake focal
zones and the distribution of focal mechanism rotations to infer
whether the CLVD component may appear as the result of fault
complexity. In Section~\ref{YYK_disc} we discuss the challenges in
studying the focal mechanism pattern and the possibility of using
local catalogs of focal mechanisms for these investigations.

\section{Tensor invariants}
\label{YYK_inv}

A seismic moment tensor can be represented as a symmetric $3 \times
3$ matrix \be {\bf m} \ = \ \left | \matrix { m_{11} & m_{12} &
m_{13} \cr m_{21} & m_{22} & m_{32} \cr m_{31} & m_{32} & m_{33} \cr
} \right | \, , \label{mom} \ee and as such it has six degrees of
freedom. The moment tensor is considered to be traceless or
deviatoric (Aki and Richards, 2002). Hence its number of degrees of
freedom is reduced to five.

The eigenvectors of matrix (\ref{mom}) are vectors
\begin{eqnarray}
{\bf t} & = & [ \, 1, \, 0, \, 0 \, ];
\nonumber\\
{\bf p} & = & [ \, 0, \, 1, \, 0 \, ];
\nonumber\\
{\bf b} & = & [ \, 0, \, 0, \, 1 \, ] \, . \label{rphi2b}
\end{eqnarray}

\begin{figure}
\begin{center}
\includegraphics[width=0.40\textwidth]{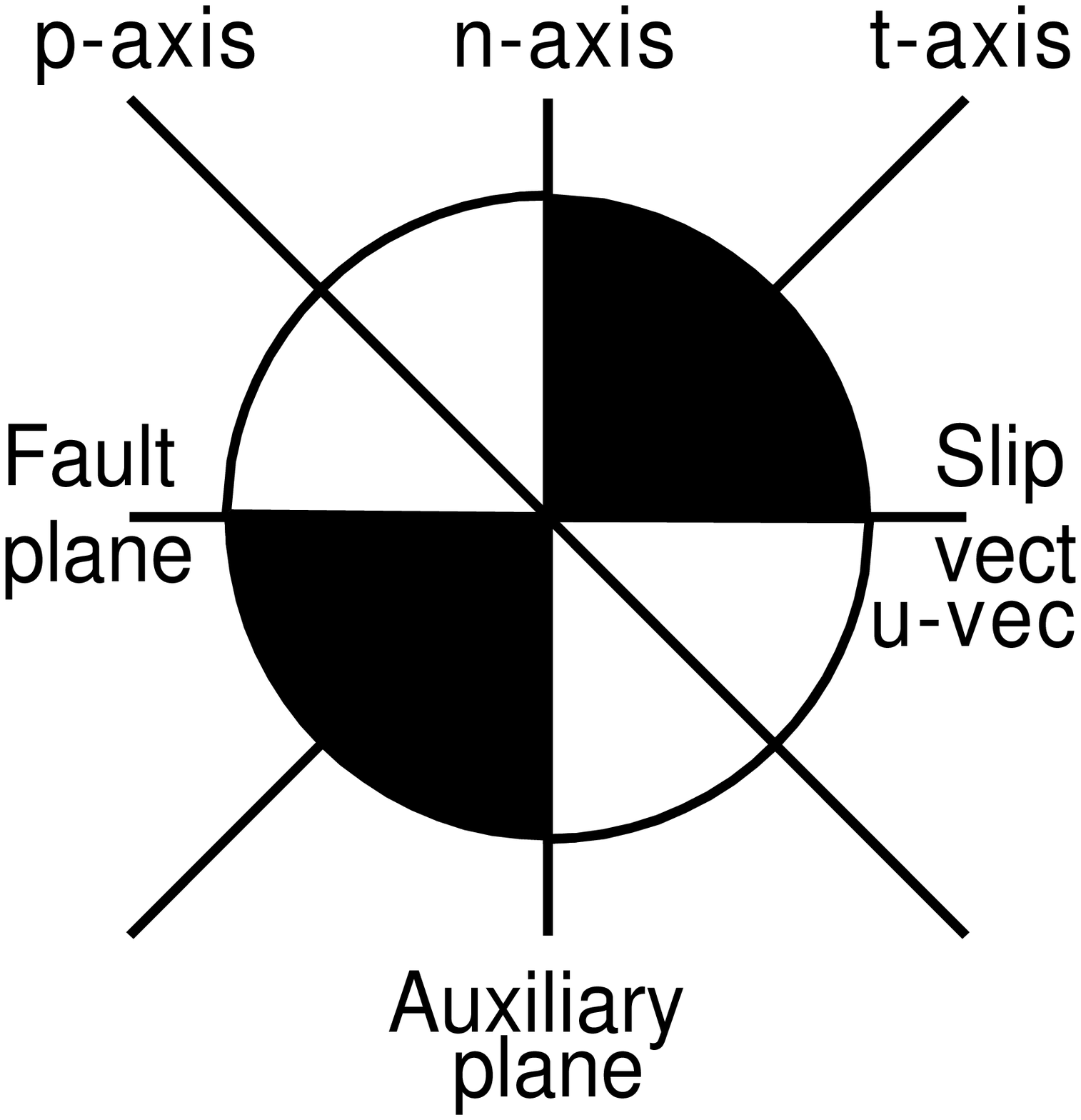}
\caption{\label{s_cal1} Schematic diagrams of earthquake focal
mechanism.
% \hfil\break
Equal-area projection (Aki and Richards, 2002, p.~110) of quadrupole
radiation patterns. The null ({\bf b}) axis is orthogonal to the
{\bf t}- and {\bf p}-axes, or it is located on the intersection of
fault and auxiliary planes, i.e., perpendicular to the paper sheet
in this display. }
\end{center}
\end{figure}

For known eigenvectors {\bf t} and {\bf p}, the DC source tensor can
be calculated as (Aki and Richards, 2002, Eq.~3.21) \ba m_{ij} \, &
= & \, \mu \, [ \, n_i u_j + n_j u_i ] \,
\nonumber \\
\, & = & \, \mu \, [ \, (t_i + p_i) \, \, (t_j - p_j) \, + (t_j +
p_j) \, \, (t_i - p_i) \, ] / 2 \, , \label{eq4a} \ea where $\mu$ is
a shear modulus, {\bf n} is a normal to a fault plane, and {\bf u}
is a slip vector (see Fig.~\ref{s_cal1}). Therefore, if we know the
orientation of two eigenvectors, the moment components can be
calculated.

The invariants of the deviatoric seismic moment tensor tensor {\bf
m} can be calculated as (Kagan and Knopoff, 1985a) \ba I_1 \, &= &
\, {\rm Tr} \, [{\bf m}] \, = \, m_{11} + m_{22} + m_{33} \,
\nonumber \\
& = & \, \lambda_1 + \lambda_2 + \lambda_3 \, \equiv \, 0 \, ,
\label{Eq_inv1} \ea where `Tr' is a trace of the tensor and
$\lambda_i$ are the eigenvalues of a moment tensor. The second
invariant or the norm of the tensor \ba I_2 \, & = & \, - \, (m_{11}
m_{22} + m_{11} m_{33} + m_{22} m_{33}) + m_{12}^2 + m_{13}^2 +
m_{23}^2 \,
\nonumber \\
&= & \, - \, (\lambda_1 \lambda_2 + \lambda_1 \lambda_3 + \lambda_2
\lambda_3) \, . \label{Eq_inv2} \ea

For a traceless tensor (\ref{Eq_inv1}) \ba I_2 \, &= & \, m_{12}^2 +
m_{13}^2 + m_{23}^2 + (m_{11}^2 + m_{22}^2 + m_{23}^2 )/2
\nonumber \\
&= & \, {1 \over 2} \sum_{i=1}^3 \sum_{j=1}^3 m_{ij}^2 \, = \,
(\lambda_1^2 + \lambda_2^2 + \lambda_3^2)/2 \, , \label{Eq_inv2a}
\ea (Jaeger and Cook, 1979, p.~33). The scalar seismic moment is \be
M \, = \, \sqrt {I_2} \, . \label{Eq_inv2b} \ee To normalize the
tensor we divide it by $M$ \be {\bf m} \, = \, {\bf m}^\prime / M \,
. \label{Eq_inv2c} \ee In the rest of the paper, unless specifically
indicated, we use only the normalized moment tensors.

The third invariant is a determinant of a tensor matrix \ba I_3 \, &
\ = \ {\rm Det} [ {\bf m } ] \, \ = \ \, m_{11} m_{22} m_{33} + 2
m_{12} m_{13} m_{23}
\nonumber \\
& - \, (m_{11} m_{23}^2 + m_{22} m_{13}^2 + m_{33} m_{12}^2 ) \ = \
\lambda_1 \lambda_2 \lambda_3 \, . \label{Eq_inv3} \ea For a
double-couple (DC) earthquake source \be I_3 \, \equiv \, 0 \, ,
\label{Eq_inv3a} \ee i.e., \be \min [ \, | \lambda_1 |, \, |
\lambda_2 |, \, | \lambda_3 | \, ] \, \equiv \, 0 \, .
\label{Eq_inv3b} \ee Thus, the normalized DC moment tensor has 3
degrees of freedom.

\section{Quaternions and representation of double-couple
earthquake mechanism and its rotation } \label{YYK_quat}

\subsection{Quaternions }
\label{YYK_quat3.1} \hfil\break
\subsubsection {Normalized quaternions }
\label{YYK_quat3.1.1} \hfil\break Kagan (1982) proposed using
normalized quaternions to represent earthquake double-couple focal
mechanism orientation. Quaternions are widely used to describe 3-D
rotations [group SO(3)] in space satellite and airplane dynamics
(Kuipers, 2002), in geodesy (Shen {\it et al.}, 2006), and in
simulations of virtual reality, robotics and automation (Kuffner,
2004; Yershova and LaValle, 2004; Hanson, 2005). There are {\scaps
matlab} packages available to perform quaternion math calculation
(MathWorks, 2006), as well as the quaternion description in {\scaps
matematica} (MathWorld, 2004).

The quaternion {\bf q} is defined as \be {\bf q} \, = \, q_0 +
q_1{\bf i} + q_2{\bf j} + q_3{\bf k} \, . \label{qu1} \ee The first
quaternion's component ($q_0$) is its scalar part, $q_1$, $q_2$, and
$q_3$ are components of a `pure' quaternion; the imaginary units
{\bf i}, {\bf j}, {\bf k}, obey the following multiplication rules
\ba & & {\bf i}^2 \, = \, {\bf j}^2 \, = \, {\bf k}^2 \, = \, -1;
\nonumber \\
& & {\bf i} \times {\bf j} \, = \, - {\bf j} \times {\bf i} \, = \,
{\bf k};
\nonumber \\
& & {\bf k} \times {\bf i} \, = \, - {\bf i} \times {\bf k} \, = \,
{\bf j};
\nonumber \\
& & {\bf j} \times {\bf k} \, = \, - {\bf k} \times {\bf j} \, = \,
{\bf i} \, . \label{qu2} \ea From (\ref{qu2}) note that the
multiplication of quaternions is not commutative, i.e., it depends
on the order of multiplicands. Non-commutability is also a property
of finite 3-D rotations. Thus, in general \be {\bf q}^{\prime
\prime} \times {\bf q}^{\prime} \ \neq \ {\bf q}^{\prime} \times
{\bf q}^{\prime \prime} \, , \label{YYK_quatpr} \ee i.e., we need to
distinguish between the right- and left-multiplication.

The conjugate ${\bf q}^*$ and inverse ${\bf q}^{-1}$ of a quaternion
are defined as \be {\bf q}^* \, = \, q_0 \, - \, q_1 \, {\bf i} \, -
\, q_2 \, {\bf j} \, - \, q_3 \, {\bf k} \quad {\rm and } \quad {\bf
q} \times {\bf q}^{-1} \, = \, 1 \, . \label{qu3} \ee

The normalized quaternion ${\bf q} = [ \, q_0, \, q_1, \, q_2, \,
q_3 \, ] $ contains four terms which can be interpreted as defining
a 3-D sphere ($S^3$) in 4-D space: \be q_0^2 \, + \, q_1^2 \, + \,
q_2^2 \, + \, q_3^2 \ = \ 1 \, . \label{eq3a} \ee Hence the total
number of degrees of freedom for the normalized quaternion is 3. For
the normalized quaternion \be {\bf q}^* = {\bf q}^{-1} \, .
\label{qu4} \ee

\hfil\break
\subsubsection {Quaternions and 3-D rotations }
\label{YYK_quat3.1.2} \hfil\break The normalized quaternion can be
used to describe a 3-D rotation: in this case the first term in
(\ref{eq3a}) represents the angle of the rotation and the following
three terms characterize the direction of its axis (Kagan, 1991).

We use normalized quaternions to calculate a rotated vector ${\cal
R}({\bf v})$ by applying the rules of quaternion multiplication
(\ref{qu2}) \be {\cal R}({\bf v}) = {\bf q} \times {\bf v} \times
{\bf q}^{-1} \, . \label{qu5} \ee The vector ${\bf v} = [q_1, q_2,
q_3]$ is represented in (\ref{qu5}) as a pure quaternion, i.e., its
scalar component is zero. In (\ref{qu5}) the quaternion {\bf q} is a
rotation operator and the pure quaternion {\bf v} is an operand
(Altmann, 1986, p.~16). Similarly to (\ref{qu5}) the whole
coordinate system can be rotated (Kuipers, 2002).

We use normalized quaternion multiplication to represent the 3-D
rotation of the DC earthquake sources. The quaternion multiplication
is \be {\bf s} \, = \, {\bf q} \times {\bf r} \, . \label{eq2a} \ee
The above expression can be written in components (Klein, 1932,
p.~61) \ba s_0 \, = \, q_0 r_0 - q_1 r_1 - q_2 r_2 - q_3 r_3 \, ;
\nonumber \\
s_1 \, = \, q_1 r_0 + q_0 r_1 \pm q_2 r_3 \mp q_3 r_2 \, ;
\nonumber \\
s_2 \, = \, \, q_2 r_0 + q_0 r_2 \pm q_3 r_1 \mp q_1 r_2 \, ;
\nonumber \\
s_3 \, = \, \, q_3 r_0 + q_0 r_3 \pm q_1 r_2 \mp q_2 r_1 \, ,
\label{eq2b} \ea where the upper sign in $\pm$ and $\mp$ is taken
for the right-multiplication and the lower sign for the
left-multiplication: ${\bf s} \, = \, {\bf r} \times {\bf q}$.

Kuipers (2002, p.~133) indicates that the right-multiplication
corresponds to the 3-D rotation of an object, whereas the
left-multiplication is the rotation of the coordinate system.
Distinguishing these multiplications is especially important when
considering a sequence of 3-D rotations \be {\bf q}^{(n)} \ = \ {\bf
q}^{\prime} \times {\bf q}^{\prime \prime} \times ... \times {\bf
q}^{(n-1)} \, , \label{YYK_quatpr1} \ee is the right-multiplication
sequence which we will use here. The corresponding rotation is
anti-clockwise with the rotation pole located on a 2-D reference
unit sphere (Altmann, 1986).

3-D rotations for quaternions of opposite signs are equal \be {\cal
S} \, [ \, {\bf q} \, ] \ = \ {\cal S} \, [ \, - {\bf q} \, ] \, ,
\label{eq3c} \ee where $ {\cal S} $ is a transformation operator of
a 3-D rotation corresponding to a quaternion {\bf q}. This means
that the group $SO(3)$ of the 3-D rotations has a two-to-one
relation to the normalized quaternions. Altmann (1986, Ch.~10)
describes the complicated topology of rotations due to this
representation.

\subsection{DC moment tensor and quaternions}
\label{YYK_quat3.2}

\hfil\break
\subsubsection {Quaternion representation of a single DC
source } \label{YYK_quat3.2.1} \hfil\break Kagan (1982) represented
the orientation of a DC source by a normalized quaternion. When
applied to the DC parametrization, the identity quaternion (zero
rotation) \be {\bf 1} = [\, 1, \, 0, \, 0, \, 0 \, ] \, ,
\label{eq3a1} \ee is identified with the strike-slip DC source with
plunge angles \be \alpha_T = \alpha_P = 0^\circ , \quad {\rm and}
\quad \alpha_B = 90^\circ \, , \label{eq3a2} \ee and azimuths \be
\beta_T = 0^\circ , \quad {\rm and} \quad \beta_P = 90^\circ \, ,
\label{eq3a3} \ee (Kagan, 1991, 2005). Any other DC source
corresponds to a quaternion describing the 3-D rotation from the
reference DC source (Eqs.~\ref{eq3a1}--\ref{eq3a3}).

There are several possible representations of rotation in 3-D. Among
the commonly used are Euler angles about coordinate axes (Kuipers,
2002, Ch.~4.3) and a rotation by the angle $\Phi$ about a rotation
axis. The rotation pole is the point where the rotation axis
intersects a reference unit sphere. We use the latter convention in
this paper since it is more convenient for the quaternion technique.
For an arbitrary DC source, the value of the rotation angle and the
spherical coordinates, $\theta$ and $\phi$, of the rotation pole on
a reference 2-D sphere ($S^2$) are then \ba \Phi & = & 2 \arccos \,
(q_0),
\nonumber \\
\theta & = & \arccos \, [q_3/\sin(\Phi/2)],
\nonumber \\
\phi & = & \arctan \, (q_2/q_1), \ {\rm if} \ \phi \le 0^\circ , \
{\rm then} \ \phi = 360^\circ + \phi \, , \label{eq3b} \ea where
$\phi$ is an azimuth ($ 360^\circ \ge \phi \ge 0^\circ$), measured
clockwise from North; and $\theta$ is a colatitude ($ 180^\circ \ge
\theta \ge 0^\circ$), $\theta = 0^\circ $ corresponds to the vector
pointing down.

We use the known correspondence between the orthogonal rotation
matrix \be {\bf R} \ = \ \left | \matrix { t_1 \ & p_1 \ & b_1 \cr
t_2 \ & p_2 \ & b_2 \cr t_3 \ & p_3 \ & b_3 \cr } \right | ,
\label{rphi4a} \ee and the normalized quaternion (Moran, 1975, Eq.\
6; Altmann, 1986, pp.\ 52, 162; Kuipers, 2002, Eq.~5.11). We obtain
the following formula for the rotation matrix \be {\bf R} \ = \
\left | \matrix { q_0^2+q_1^2-q_2^2-q_3^2 \ & 2 \, (-q_0q_3 + q_1q_2
) \ & 2 \, (q_0q_2+ q_1q_3 ) \cr 2 \, (q_0q_3 + q_1q_2 ) \ &
q_0^2-q_1^2+q_2^2-q_3^2 \ & 2 \, (-q_0q_1 + q_2q_3 ) \cr 2 \,
(-q_0q_2 + q_1q_3 ) \ & 2 \, (q_0q_1 + q_2q_3 ) \ &
q_0^2-q_1^2-q_2^2+q_3^2 \cr } \right | , \label{rphi5} \ee to obtain
the quaternion's components. The above formula can be obtained by
applying (\ref{qu5}) to each of the original {\bf t}, {\bf p}, and
{\bf b} vectors (\ref{rphi2b}). Kagan and Knopoff (1985, their
Eq.~5) provide another expression for the rotation matrix using
direction cosines of the axes [the first term in the second matrix
row should be corrected as $ \ell m(1 -\cos\Phi ) - m\sin\Phi $].

Comparing (\ref{rphi4a}) with (\ref{rphi5}) we derive quaternion
components from the rotation matrix direction cosines (Kuipers,
2002, p.~169; Hanson, 2005, pp.~149-150). For example, if $q_0$ is
not close to zero \ba q_0 \ & = & \ (1/2) \, \sqrt { t_1 + p_2 + b_3
+ 1 } \, ;
\nonumber \\
q_1 \ & = & \ ( b_2 - p_3 ) /(4 q_0) \, ;
\nonumber \\
q_2 \ & = & \ ( t_3 - b_1 ) /(4 q_0) \, ;
\nonumber \\
q_3 \ & = & \ ( p_1 - t_2) /(4 q_0) \, , \label{rphi5f} \ea Since as
many as three of the quaternion components may be close to zero, it
is computationally simpler to use the component with a maximum
absolute value to calculate the three other components.

The DC seismic moment tensor in eigenvector coordinates is ${\bf m}
= {\rm diag} [\, 1, \, -1, \, 0 \, ]$. For the general orientation
of an earthquake focal mechanism, the seismic moment tensor
(\ref{mom}) can be calculated from the normalized quaternion as
follows (Kagan and Jackson, 1994):

\begin{eqnarray}
m_{11} & = & q_1^4-6q_1^2q_2^2 - 2q_1^2q_3^2+2q_1^2q_0^2 +
8q_1q_2q_3q_0+q_2^4
\nonumber\\
& & + \, 2q_2^2q_3^2-2q_2^2q_0^2 + q_3^4-6q_3^2q_0^2+q_0^4 \, ;
\nonumber\\
m_{12} & = & 4 \, (q_1^3q_2 - q_1q_2^3 - q_3^3q_0 + q_3q_0^3) \, ;
\nonumber\\
m_{13} & = & 2 \, (q_1^3q_3 - 3 q_1^2q_2q_0 - 3q_1q_2^2q_3 -
q_1q_3^3
\nonumber\\
& & + \, 3q_1q_3q_0^2 + q_2^3q_0 + 3q_2q_3^2q_0 - q_2q_0^3) \, ;
\nonumber\\
m_{22} & = & - \, q_1^4+6q_1^2q_2^2 - 2q_1^2q_3^2+2q_1^2q_0^2 +
8q_1q_2q_3q_0
\nonumber\\
& & - \, q_2^4 + 2q_2^2q_3^2-2q_2^2q_0^2 - q_3^4+6q_3^2q_0^2-q_0^4
\, ;
\nonumber\\
m_{23} & = & 2 \, (q_1^3q_0 + 3q_1^2q_2q_3 - 3q_1q_2^2q_0 +
3q_1q_3^2q_0
\nonumber\\
& & - \, q_1q_0^3 - q_2^3q_3 + q_2q_3^3 - 3q_2q_3q_0^2) \, ;
\nonumber\\
m_{33} & = & 4 \, (q_1^2q_3^2 - q_1^2q_0^2 - 4q_1q_2q_3q_0 -
q_2^2q_3^2 + q_2^2q_0^2) \, . \label{rphi2}
\end{eqnarray}

\hfil\break
\subsubsection {Mutual rotation of DC sources }
\label{YYK_quat3.2.2} \hfil\break A more complicated algorithm is
needed for the rotation of any DC source into another. The methods
of quaternion algebra can be used to evaluate the 3-D rotation angle
by which one DC source can be so transformed (Kagan, 1991).
Alternatively, the standard technique of orthogonal matrices can be
applied to this calculation (Kagan, 2007b).

Given the symmetry of the DC source (Kagan and Knopoff, 1985a;
Kagan, 1990; 1991) the $q_0$ term in (\ref{eq3a}) can always be
presented as the largest positive term in this parameterization. In
particular, to obtain the standard DC quaternion representation, we
right-multiply an arbitrary normalized quaternion $ {\bf q}$ by one
of the elementary quaternions (Kagan, 1991): \ba {\bf i} = [\, 0, \,
1, \, 0 \, ,0 \, ] \, ;
\nonumber \\
{\bf j} = [\, 0, \, 0, \, 1, \, 0 \, ] \, ;
\nonumber \\
{\bf k} = [\, 0, \, 0, \, 0, \, 1 \, ] \, , \label{eq3d} \ea if the
second, third or fourth term has the largest absolute value,
respectively. For example, for the largest second term, $q_1$ \be
{\bf q}^{\prime \prime} \ = \ {\bf q}^\prime \, \times \, {\bf i} \,
. \label{eq3e} \ee If the resulting first term is negative, the sign
of all terms should be reversed (see Eq.~\ref{eq3c}).

As the result of multiplication by expressions (\ref{eq3d}) the
quaternion $\bf q$ becomes \ba {\bf q} & \times & {\bf 1} \, = \, [
\, q_0, \, q_1, \, \, q_2, \, \, q_3 \, ];
\nonumber \\
{\bf q} & \times & {\bf i} \, = \, [ \, q_1, - q_0, \, - q_3, \, q_2
\, ];
\nonumber \\
{\bf q} & \times & {\bf j} \, = \, [ \, q_2, \, q_3, \, - q_0, \, -
q_1 \, ];
\nonumber \\
{\bf q} & \times & {\bf k} \, = \, [ \, q_3, - q_2, \, q_1, \, - q_0
\, ] \, . \label{eq3f} \ea The transformations (\ref{eq3c},
\ref{eq3f}) describe an eight-to-one correspondence between an
arbitrary normalized quaternion and a quaternion corresponding to a
normalized seismic moment tensor. We call this operator ${\cal Q}$.
%Therefore, the quaternion {\bgr \char22} \be {\bfg \gkm} \ = \ {\cal
%Q} \, ( {\bf q} ) \, , \label{eq3g} \ee is a one-to-one quaternion
%representation of a DC focal mechanism. We call this operator ${\cal
%Q}$.
Therefore, the quaternion {$\xi$} \be {\xi} \ = \ {\cal Q} \, ( {\bf
q} ) \, , \label{eq3g} \ee is a one-to-one quaternion representation
of a DC focal mechanism. We call this operator ${\cal Q}$.

It is easy to check that all the eight quaternions of the ${\cal Q}$
operator (\ref{eq3g}) produce the same moment tensor (\ref{rphi2}).
Thus we can write \be {\bf m} \, = \, {\cal M} \, ( {\bf q} ) \, =
\, {\cal M} \, ( {\xi} ) \, , \label{rphi2a} \ee where ${\cal M} \,
( \, )$ is an operator (\ref{rphi2}) converting a quaternion into a
seismic moment tensor matrix.

In principle, we can use the non-normalized variables. In this case
the norm of a quaternion would correspond to that of the tensor
(i.e., a scalar seismic moment $M$ for a DC source, see
Eq.~\ref{Eq_inv2b}). However, the general deviatoric tensor
(Eq.~\ref{Eq_inv1}, see also Section~\ref{YYK_gam}) has five degrees
of freedom even after it has been normalized. Hence it cannot be
represented by a regular quaternion.

Similarly to (\ref{qu5}) rotation of a DC requires quaternion
multiplication (as shown, for example in Eq.~\ref{YYK_quatpr1}). The
rotated DC source then needs to be converted into a DC standard
quaternion representation using (\ref{eq3g}).

Thus, in our representation, an arbitrary quaternion is both a
rotation operator and a DC source after simple transformations
(\ref{eq3g}) have been performed (Kagan, 2005). Although the
quaternion does not have the advantage of clearly identifying the DC
source properties, its benefits are obvious. Multiple rotations of
the DC source as well as the inverse problem determining the
rotation from one source to another are easily computed using the
methods of quaternion algebra (Kagan, 1991; Ward, 1997; Kuipers,
2002).

The {\scaps fortran} program which determines the 3-D rotation of DC
sources is available on the Web -- \hfil\break
ftp://minotaur.ess.ucla.edu/pub/kagan/dcrot.for (see also {\scaps
fortran90} adaptation of the programme by P.~Bird
http://peterbird.name/oldFTP/2003107-esupp/Quaternion.f90.txt).
Frohlich and Davis (1999) also discuss the program. Kagan (2007b)
supplies simplified algorithms for calculating a 3-D DC rotation.
These algorithms can be written in a few lines of computer code.

\subsection{Lambert azimuthal equal-area projection }
\label{YYK_quat3.3}

The DC focal mechanism has a symmetry of a rectangular box (Kagan,
1991; 2005). Thus, it is convenient to use the Lambert azimuthal
equal-area projection of an octant for a display of many
distributions associated with the DC source. The coordinates of the
projection are (Snyder, 1987, p.~185) \ba X \ &=& \ C \, \cos
(\theta^\prime) \, \sin (\, \phi^\prime - \phi^\prime_0 \, );
\nonumber \\
Y \ &=& \ C \, \left [ \, \cos (\theta^\prime_0) \, \sin
(\theta^\prime) \, - \, \sin (\theta^\prime_0) \, \cos
(\theta^\prime) \, \cos (\, \phi^\prime \, - \, \phi^\prime_0 \, )
\, \right ] ;
\nonumber \\
C \ &=& \ \sqrt { { 2 \over { 1 + \sin (\theta^\prime_0) \, \sin
(\theta^\prime) \, + \, \cos (\theta^\prime_0) \, \cos
(\theta^\prime) \, \cos (\phi^\prime - \phi^\prime_0) } } } \, ,
\nonumber \\
\label{Eq_g8} \ea where $180^\circ \ge \phi^\prime \ge - 180^\circ $
is the centered longitude ({\it cf.}~Eq.~\ref{eq3b}), $90^\circ \ge
\theta^\prime \ge - 90^\circ $ is the latitude; $\phi^\prime_0$ and
$\theta^\prime_0$ are the coordinates of the projection center. For
octant projection we use $\phi^\prime_0 = 45^\circ$ and
$\theta^\prime_0 = \arctan {1 \over \sqrt 2 } \ \approx \
35.26^\circ$. Then (\ref{Eq_g8}) can be simplified \ba X \ &=& \ C
\, \cos (\theta^\prime) \, \sin (\, \phi^{\prime\prime}) ;
\nonumber \\
Y \ &=& \ {C \, \over \sqrt 3 } \left [ \, \sin (\theta^\prime) \, -
\, \cos (\theta^\prime) \, \cos (\phi^{\prime\prime}) \, \right ] ;
\nonumber \\
C \ &=& \ \sqrt { { { 2 \, \sqrt 3 } \over { \sqrt 3 + \sin
(\theta^\prime) \, + \, \sqrt 2 \, \cos (\theta^\prime) \, \cos
(\phi^{\prime\prime}) } } } \, , \label{Eq_g8a} \ea where
$\phi^{\prime\prime} = \phi^\prime - 45^\circ$. Kagan (2005,
Eqs.~26-29) provides an equivalent equal-area projection formula for
an octant, if plunge angles $\alpha_i$ (\ref{eq3a2}) of a DC
solution are known.

\section{CLVD and Gamma index }
\label{YYK_gam}

The Gamma index (Kagan and Knopoff, 1985a; Frohlich, 1990;
Richardson and Jordan, 2002) is \be \Gamma \ = \ { { 3 \sqrt 3 }
\over 2 } \times { { I_3 } \over I_2^{3/2} } \, , \label{Eq_gam} \ee
(see Eqs.~\ref{Eq_inv2}--\ref{Eq_inv3a}). For a DC source \be \Gamma
\ \equiv \ 0 \, , \label{Eq_gam1} \ee (Eq.~\ref{Eq_inv3a}). The
$\Gamma$-index ranges from $-1$ to 1; $ | \Gamma | = 1$ corresponds
to a pure CLVD source (Knopoff and Randall, 1970; Kagan and Knopoff,
1985a).

\subsection{Gamma index for purely random rotation }
\label{YYK_gam1}

Kagan and Knopoff (1985a) considered the problem of a CLVD index
distribution for a composite source \be {\bf m} \, = \, \sum_{i=1}^N
\, {\bf R}_i \, {\bf m}^{(i)} \, {\bf R}_i^T \, , \label{Eq_gam2}
\ee where ${\bf R}_i$ is a random rotation matrix and ${\bf R}_i^T$
is its transpose. They showed that for the large number of summands
($N$) the resulting source has the uniform $\Gamma$-index
distribution in the range $1.0 \ge \Gamma \ge -1.0 $. This result
may be used to explain the non-zero $\Gamma$-index value, sometimes
obtained for earthquakes with a complex fault zone, i.e., an
earthquake source comprising several DC components of different
orientation. However, there are both theoretical and observational
arguments suggesting that the structure of a source is complex for
tectonic earthquakes but precludes the appearance of a CLVD
component.

In a quaternion notation, (\ref{Eq_gam2}) can be expressed as \be
{\cal M} \, ( {\bf q} ) \, = \, \sum_{i=1}^N {\cal M} \, ( {\bf
q}^{(i)} \times {\bf r} ) \, , \label{Eq_gam3} \ee where the
operator ${\cal M}$ is given by (\ref{rphi2a}) and ${\bf r}$ is a
random rotation quaternion that can be obtained using Marsaglia's
(1972) algorithm (see more in Kagan, 2005).

Kagan and Knopoff (1985a, their Fig.~1a) used simulation to show
that for the sum of randomly oriented DC sources the $\Gamma$-index
is distributed uniformly over the interval [$-1$,~1] for a large
number of summands. Even for two DCs, the distribution is close to
uniform. This simplicity of the $\Gamma$-index distribution presents
a significant benefit in characterizing the CLVD component. Many
other measures of non-DC properties for an earthquake source have
been proposed (see, for example, Julian {\it et al.}\ 1998, Eq.~18),
but lack our statistical advantage.

\subsection{Gamma index for composite sources }
\label{YYK_gam2}

\hfil\break
\subsubsection{General considerations }
\label{YYK_gam2.1} \hfil\break

We will now explore the $\Gamma$-index distribution when the 3-D
rotation is not completely random, i.e., if the rotation pole is
preferentially located in a DC focal mechanism. For simplicity we
assume that only two DCs comprise the composite source ($N=2$ in
Eq.~\ref{Eq_gam3}) \be {\cal M} \, ( {\bf s} ) \, = \, {\bf q} \, +
\, \chi \, {\cal M} \, ( {\bf q} \times {\bf r} ) \, ,
\label{Eq_gam4} \ee where for generality we assume that these DC
sources have different scalar moments and their moment ratio is
$\chi = M^{\prime} \, / \, M^{\prime \prime} $. Since the quaternion
{\bf q} is arbitrary, we can select it to be the identity
(\ref{eq3a1}). The general rotation quaternion is (see
Eq.~\ref{eq3b}) \ba r_0 & = & \sqrt {1 - A^2} \, ;
\nonumber \\
r_1 & = & A \, \sin(\theta) \, \cos(\phi) \, ;
\nonumber \\
r_2 & = & A \, \sin(\theta) \, \sin(\phi) \, ;
\nonumber \\
r_3 & = & A \, \cos(\theta) \, , \label{Eq_gam5} \ea where \be A =
\sin(\Phi/2) \, , \label{Eq_gam5b} \ee (see Eq.~\ref{eq3b}).

Using {\scaps mathematica} (Wolfram, 1999), we calculate the third
invariant (\ref{Eq_inv3}) for (\ref{Eq_gam4}). We find that if the
rotation axis is ${\bf b}$, i.e., $\theta = 0^\circ$ in
(\ref{Eq_gam5}), $I_3 \equiv 0$ for all $\Phi$'s. Similarly, if the
rotation axis is either the normal to the fault plane or is a slip
vector \be \phi = 45^\circ \quad {\rm or} \quad \phi = 135^\circ
\quad {\rm and} \quad \theta = 90^\circ \, , \label{Eq_gam5a} \ee
the invariant is also zero for all $\Phi$'s. This means that the sum
of two DC sources is again a DC for these rotations (Frohlich {\it
et al.}, 1989). The same result ($I_3 \equiv 0$) is obtained, if
more than two DC sources are rotated around the {\bf b}-axis or the
{\bf n}- and {\bf u}-axes (see Eq.~\ref{eq4a}) and then added with
different weights (as in Eq.~\ref{Eq_gam4}).

\begin{figure}
\begin{center}
\includegraphics[width=0.45\textwidth,angle=0]{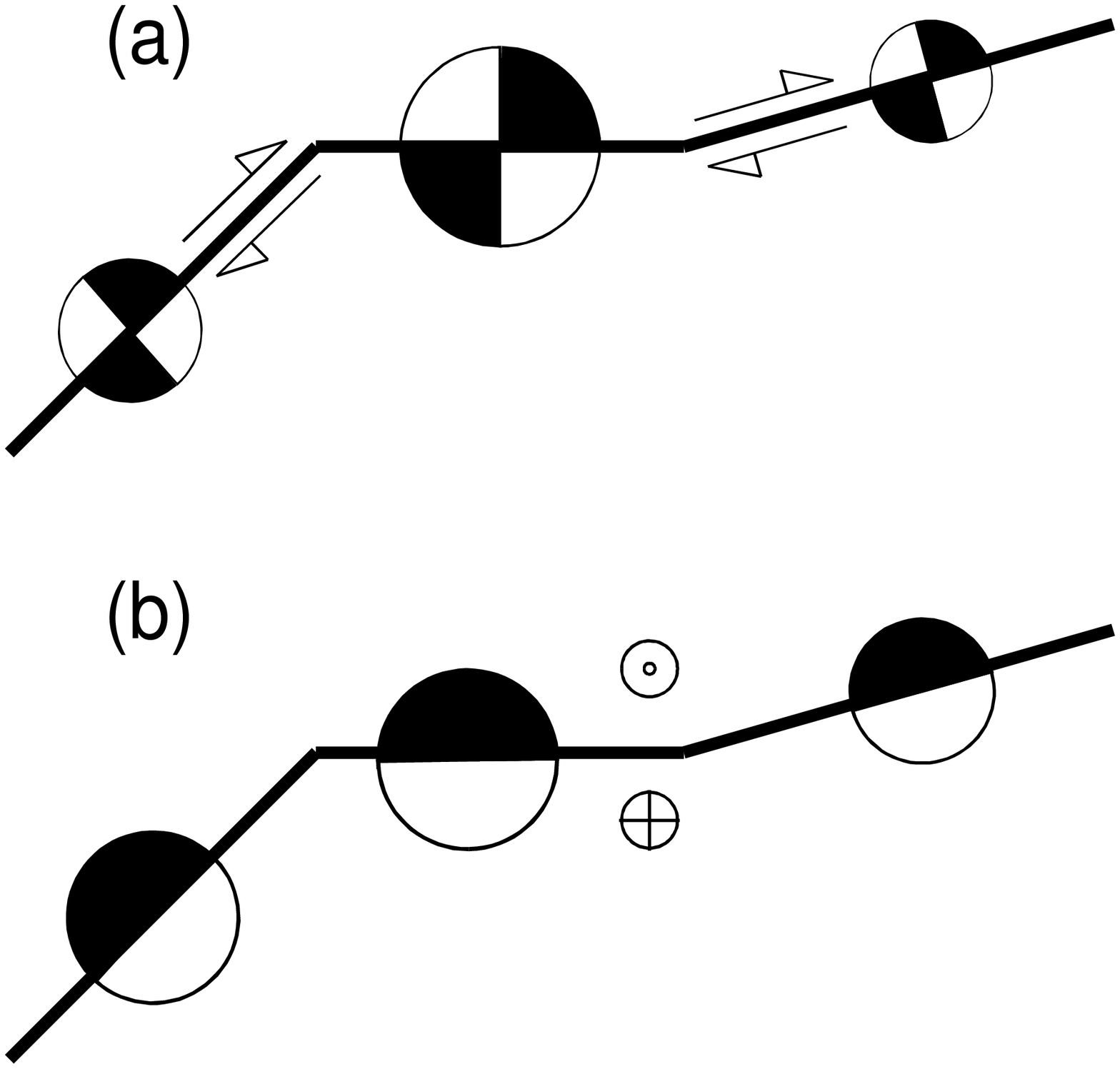}
\caption{\label{s_cal3a} Schematic diagrams of earthquake focal
zone. Lines show a complex fault plane; focal mechanisms of
earthquake(s) are displayed for each fault segments. \hfil\break (a)
Rotation around {\bf b}-axis, non-conservative geometric barriers.
\hfil\break (b) Rotation around {\bf u}-axis (slip vector),
conservative geometric barriers. }
\end{center}
\end{figure}

\begin{figure}
\begin{center}
\includegraphics[width=0.45\textwidth,angle=0]{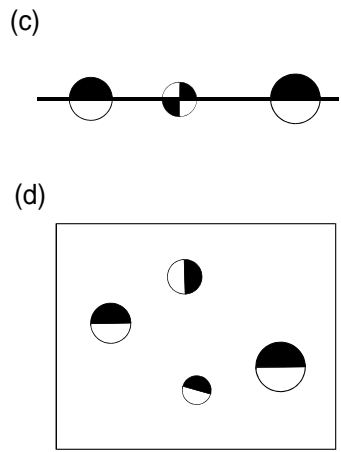}
\caption{\label{s_cal3b} Schematic diagrams of earthquake focal
zone. Rotation around {\bf n}-axis (fault-plane normal vector),
conservative geometric barrier. \hfil\break (c) View from side.
\hfil\break (d) View from above. }
\end{center}
\end{figure}

King (1986) specified two kinds of geometric barriers connected with
a change of earthquake failure surfaces: conservative and
non-conservative. The former structure does not require creating new
faulting or void space. In our notation it would correspond to the
second case (\ref{Eq_gam5a}). The first case is a non-conservative
system: the incompatibility of displacement would require producing
new ruptures (Gabrielov {\it et al.}, 1996). Therefore, with both
kinds of barriers a complex geometric source is still a DC.
Figs.~\ref{s_cal3a} and \ref{s_cal3b} display cartoons of possible
fault-plane and focal mechanism arrangements for these barriers.

\hfil\break
\subsubsection{Small rotations }
\label{YYK_gam2.2} \hfil\break

\begin{figure}
\begin{center}
\includegraphics[width=0.60\textwidth,angle=0]{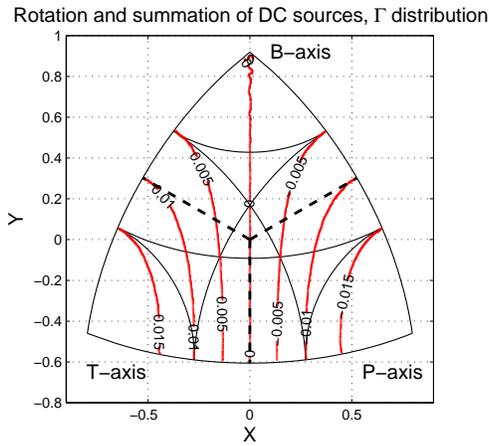}
% Fig.4, P43, p.46
\caption{\label{s_cal4} $\Gamma$-index distribution for small
rotation angles ($\Phi=10^\circ$). See Eq.~\ref{Eq_gam6}. Two equal
DC sources, one rotated compared to another. The axis angles are
shown at octant equal-area projection (\ref{Eq_g8a}); see also
(Kagan, 2005). Dashed lines are boundaries between different focal
mechanisms. Plunge angles $30^\circ$ and $60^\circ$ for all
mechanisms are shown by thin solid lines. }
\end{center}
\end{figure}

If $A$ in (\ref{Eq_gam5b}) is small, we can keep only lower order
terms in (\ref{Eq_gam4}) and obtain for the third invariant
(\ref{Eq_inv3}) of the sum \ba I_3 ( {\bf s} ) \, & \approx & \, 4
\, A^2 \, \chi \, (1 + \chi) \, \sin^2(\theta) \, \left [
\cos^2(\phi) - \sin^2(\phi) \right ]
\nonumber \\
& + & 16 \, A^3 \, \sqrt {1 - A^2} \, \chi \, (1 - \chi)
\nonumber \\
& \times & \sin^2(\theta) \, \cos(\theta) \, \sin(\phi) \,
\cos(\phi) \, . \label{Eq_gam6} \ea The first term in the right-hand
part of the equation suggests that the invariant reaches maximum in
the equatorial plane ($\theta~=~90^\circ$) when the pole coincides
with the {\bf t}- or {\bf p}-axes ($\phi~=~0^\circ$ or
$\phi~=~90^\circ$). Fig.~\ref{s_cal4} displays the distribution of
the $\Gamma$-index for the value of the rotation angle
$\Phi=10^\circ$ and for equal DC components ($\chi=1.0$).

In calculations we use an approximate formula (\ref{Eq_gam6}); the
exact expression (obtained by {\scaps mathematica}), which is too
long to show here, yields almost the same answer. The $\Gamma$-index
values in Fig.~\ref{s_cal4} are very small; thus, a non-zero CLVD
component is unlikely to be obtained in the moment solutions for an
earthquake consisting of such subevents.

\hfil\break
\subsubsection{Large rotations }
\label{YYK_gam2.3} \hfil\break

\begin{figure}
\begin{center}
\includegraphics[width=0.60\textwidth,angle=0]{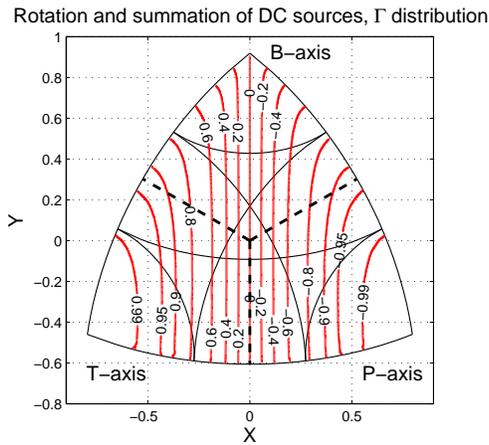}
% Fig.5, P43, p.46
\caption{\label{s_cal5} $\Gamma$-index distribution for large
rotation angles ($\Phi=90^\circ$). Two equal DC sources, one rotated
compared to another. Octant projection and auxiliary lines are the
same as in Fig.~\ref{s_cal4}. }
\end{center}
\end{figure}

Formula (\ref{Eq_gam6}) also suggests that if $\phi = 45^\circ$, for
small values of the rotation angle $\Phi$ and for non-equal DC
components the third invariant is proportional to $\sin^3(\Phi/2)$,
i.e., it is close to zero. In particular, if $\chi =1$, by employing
$\phi = 45^\circ$ in (\ref{Eq_gam6}), we obtain that $I_3 \equiv 0$
for arbitrary values of the rotation angle $\Phi$. This means that
if the rotation pole is located on a fault- or auxiliary-plane, the
sum of two focal mechanisms (original and rotated) has a zero CLVD
component, i.e., it is a pure DC source. The effect can be seen in
Fig.~\ref{s_cal5}, where the $\Gamma$-index, calculated with the
exact formula, is displayed for equal DC components. In this case,
if the rotation pole is close to the {\bf t}- or {\bf p}-axis, the
resulting source is almost pure CLVD.

This property can be demonstrated for a few simple arrangements of
sources (see also Frohlich {\it et al.}, 1989). For example, a sum
of two DCs \be {\bf s} \ = \ \left | \begin{array}{rrr}
1 & 0 & 0 \\
0 & -1 & 0 \\
0 & 0 & 0
\end{array}
\right | \ + \ \left | \begin{array}{rrr}
0 & 1 & 0 \\
1 & 0 & 0 \\
0 & 0 & 0
\end{array}
\right | \, , \label{Eq_gam7} \ee can be rotated into ${\bf s} \, =
\, {\rm diag} [\sqrt{2}, \, - \sqrt{2}, \, 0]$ with
$I_3(s)~\equiv~0$. This arrangement can be represented as two
strike-slip events with a fault-plane rotated by $45^\circ$.

Another example is the rotation by $120^\circ$ around a pole with
coordinates $[1/\sqrt{3}, \, 1/\sqrt{3}, \, 1/\sqrt{3}]$, a turn
which exchanges the position of coordinate axes. \be {\bf s} \ = \
{\rm diag} [1, \, - 1, \, 0] \ + \ {\rm diag} [-1, \, 0, \, 1] \ = \
{\rm diag} [0, \, -1, \, 1] \, . \label{Eq_gam8} \ee Again $I_3(s)
\equiv 0$.

For rotation around {\bf t}-axis, for example, \be {\bf s} \ = \
{\rm diag} [1, \, - 1, \, 0] \ + \ {\rm diag} [1, \, 0, \, -1] \ = \
{\rm diag} [2, \, -1, \, -1] \, . \label{Eq_gam9} \ee Such a sum
would have $\Gamma = 1$, i.e., it is a pure CLVD source (see
Fig.~\ref{s_cal5}).

The results above demonstrate that if the rotation pole is located
even randomly on nodal-planes, the resulting sum source is a DC.
This may be relevant in searching for a non-DC component in seismic
moment solutions. Frohlich and Davis (1999) and Kagan (2003) argued
that the CLVD component is not presently measured with accuracy
sufficient to study its properties.

\section{Tensor moment dot product }
\label{YYK_mdot}

The tensor dot product has been introduced by Kagan and Knopoff
(1985b) as one of the correlation tensor invariants describing a
complex earthquake fault pattern. Alberti (2006) used it to
characterize the similarity of earthquake focal mechanisms. We
compare two methods for characterizing earthquake source complexity:
rotation angle $\Phi$ and tensor dot product \be D \ = \ m_{ij} \,
n_{ij}; \qquad 2 \, \ge \, D \, \ge \, -2 \, , \label{Eq_gam6d} \ee
where the second source is rotated with regard to the first by the
angle $\Phi$.

The dependence of the product $D$ on 3 parameters of 3-D rotation
(\ref{eq3b}) calculated using {\scaps mathematica}, can be described
as \ba D \ = & \ 2 \, \bigg \{ \, (\, 1 - \,{{A}^2} \, )^2 -
\,{{A}^2}\,{{\cos^2 (\theta)}}
\nonumber \\
& \times \Big [\, -6 \, + \, 6\,{{A}^2}\, + \,{{A}^2}\,{{\cos^2
(\theta)}} \Big ] + \,{{A}^4}\,{{\sin^4 (\theta)}}
\nonumber \\
& \times \Big [ {{\cos^4 (\phi)}}\, - 6\,{{\cos^2 (\phi)}}\,{{\sin^2
(\phi)}} \, + \,{{\sin^4 (\phi)}}\, \Big ] \bigg \} \, ,
\label{Eq_gam6c} \ea where $A$ is given by (\ref{Eq_gam5b}).

\begin{figure}
\begin{center}
\includegraphics[width=0.60\textwidth,angle=0]{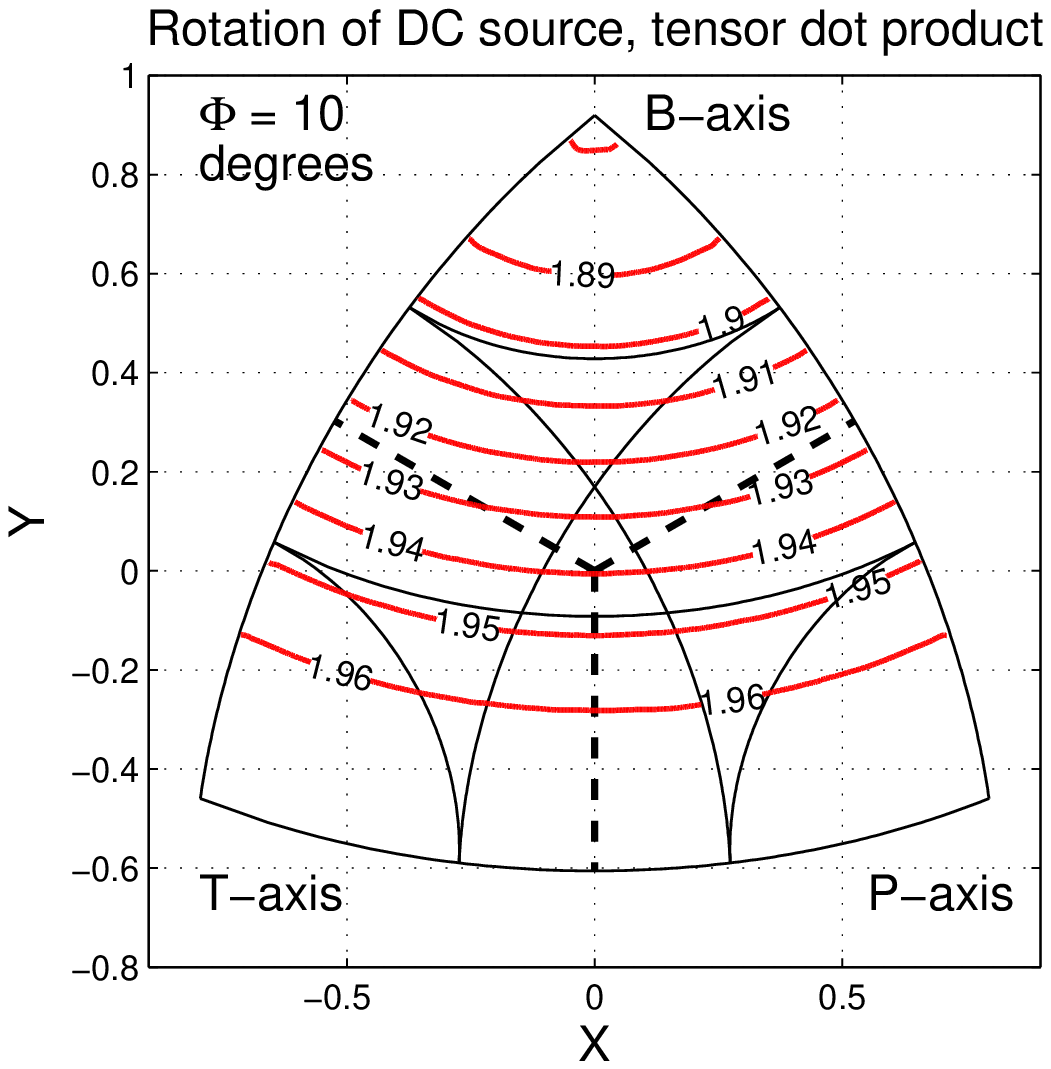}
\caption{\label{s_cal6a} Distribution of the tensor dot product for
rotation angle $\Phi=10^\circ$. Two equal DC sources, one rotated
compared to another. Octant projection and auxiliary lines are the
same as in Fig.~\ref{s_cal4}. }
\end{center}
\end{figure}

\begin{figure}
\begin{center}
\includegraphics[width=0.60\textwidth,angle=0]{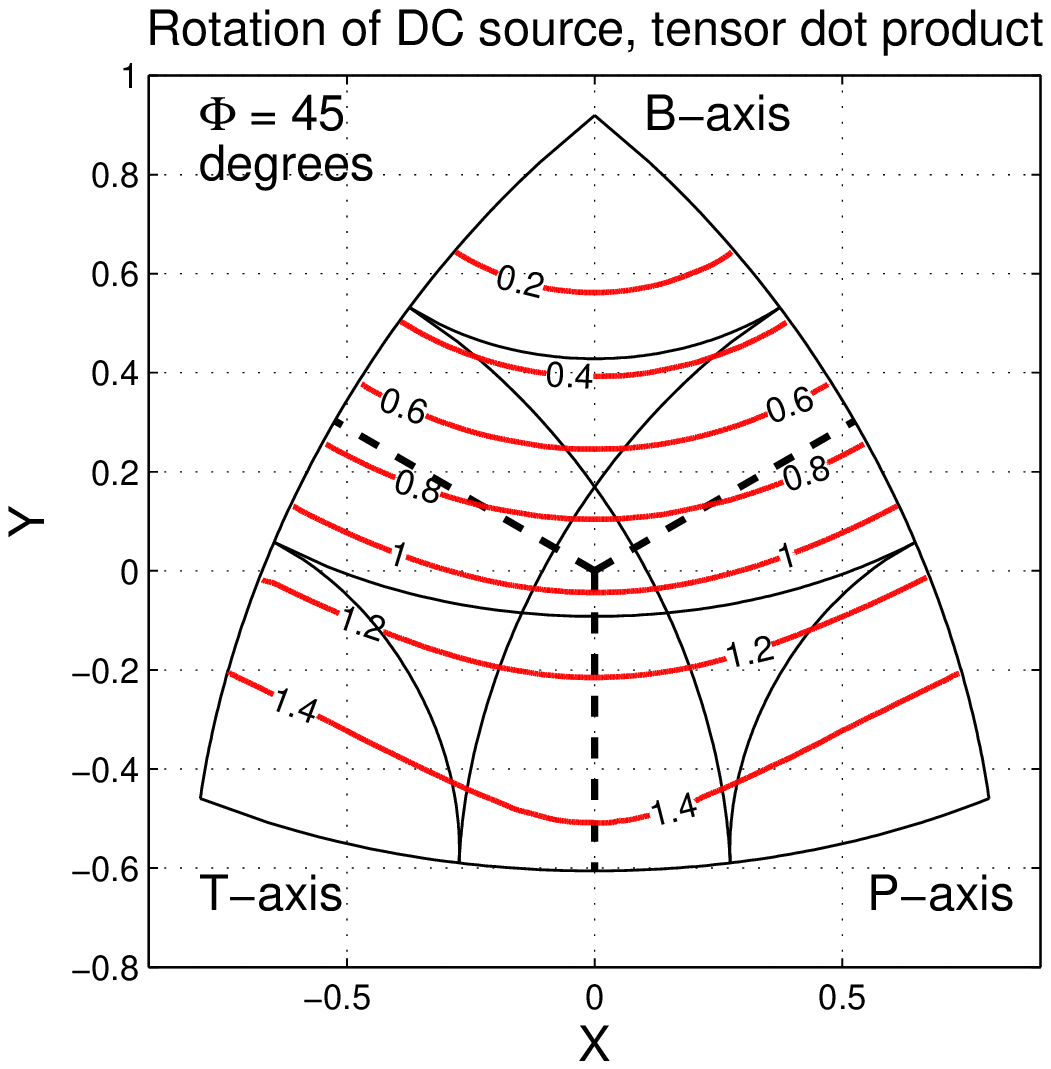}
\caption{\label{s_cal6b} Distribution of the tensor dot product for
rotation angle $\Phi=45^\circ$. Two equal DC sources, one rotated
compared to another. Octant projection and auxiliary lines are the
same as in Fig.~\ref{s_cal4}. }
\end{center}
\end{figure}

\begin{figure}
\begin{center}
\includegraphics[width=0.60\textwidth,angle=0]{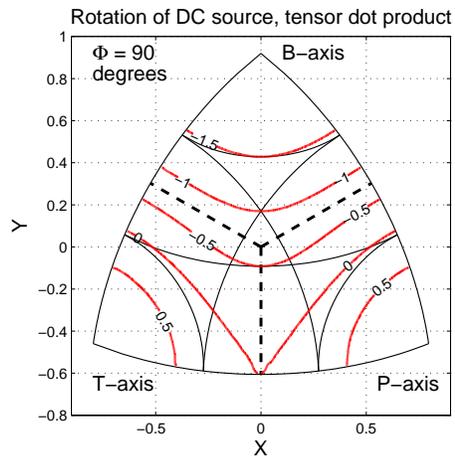}
\caption{\label{s_cal6c} Distribution of the tensor dot product for
rotation angle $\Phi=90^\circ$. Two equal DC sources, one rotated
compared to another. Octant projection and auxiliary lines are the
same as in Fig.~\ref{s_cal4}. The $D$-value is $-2$ at the {\bf
b}-axis and $+1$ at the {\bf t}- and {\bf p}-axes. }
\end{center}
\end{figure}

In Figs.~\ref{s_cal6a}--\ref{s_cal6c} we display the $D$-value for
three choices of the $\Phi$ angle. The maximum change of the
$D$-value from $D=2$, corresponding to identical DC sources, occurs
when the rotation pole is at the {\bf b}-axis. For $\Phi= 90^\circ$
rotation $D=-2$ at the {\bf b}-axis and $D=1$ for other axes ({\it
cf.}~Eq.~\ref{Eq_gam9}). Hence, the range of the $D$ change is four
times greater for the {\bf b}-axis rotation than for rotation at
other axes. Therefore, from the $D$-value alone we cannot fully
infer the complexity and coherence of DC sources. Higher order
invariants of the correlation tensor need to be studied to better
describe the fault pattern complexity (Kagan, 1992a).

\section{Focal mechanisms statistics }
\label{YYK_stat}

The orientation of a DC source can be characterized by the following
three quantities: a rotation angle ($\Phi$) of the counterclockwise
rotation from the first DC source to the second, and a location of a
rotation pole on a reference sphere (\ref{eq3b}) -- colatitude,
$\theta$, and longitude, $\phi$ (Kagan 1991, 2003). Thus, to fully
study the distribution of earthquake focal mechanisms we need to
investigate a six-dimensional manifold (fiber bundle): a product of
the 3-D Euclidean space and the 3-D rotational distribution of a DC
source. This presents a double difficulty~-- there is no effective
procedure to display and study this pattern and we do not have
sufficient data to evaluate the parameters of the six-dimensional
distribution.

Therefore, to better analyze the focal mechanism pattern, we should
construct only marginal distributions of the full 6-D manifold. The
following examples serve to illustrate the theoretical methods
described in previous sections.

In Section~\ref{YYK_stat2} we analyze the distribution of earthquake
centroids projected on a focal mechanism reference sphere for each
earthquake. Section~\ref{YYK_stat3} presents the distribution of
rotation angles between pairs of DC solutions, and
Section~\ref{YYK_stat4} discusses the statistics of rotation poles.

\subsection{Earthquake catalog }
\label{YYK_stat1}

We study the earthquake distributions for the global catalog of
moment tensor inversions compiled by the CMT group (Ekstr\"om {\it
et al.}, 2005, and references therein; see also references and
earthquake statistics for 1977-1998 in Dziewonski {\it et al.},
1999). The catalog contains 26,865 solutions over a period from
1977/1/1 to 2007/3/31. Only shallow earthquakes (depth 0-70~km) are
studied here.

The CMT catalog includes seismic moment centroid times and locations
as well as estimates of seismic moment tensor components (Dziewonski
{\it et al.}, 1981; Dziewonski and Woodhouse, 1983). Each tensor is
constrained to have zero trace (first invariant), i.e., no isotropic
component. Double-couple (DC) solutions, i.e., with tensor
determinant equal to zero, are supplied as well. Almost all
earthquake parameters are accompanied by internal estimates of
error.

From the original CMT catalog we created subcatalogs of
well-constrained solutions. These datasets are obtained by removing
from the catalog (a) the earthquakes lacking all 6 independent
components of the moment tensor, (b) solutions with a large relative
error, and (c) the solutions with large CLVD component (Frohlich and
Davis, 1999; Kagan, 2000). Since the larger earthquakes usually have
smaller errors (Kagan, 2003), fewer of these events are removed.
About 85\% of $m \ge 6.0$ earthquakes are well-constrained, whereas
for smaller earthquakes ($m \ge 5.0$) more than 2/3 have been
deleted.

\subsection{Centroid distribution on a focal sphere }
\label{YYK_stat2}

Similarly to Table~1 in Kagan (1992b) in Table~1 here we show the
distribution of numbers of centroids in a coordinate system formed
by the {\bf t}-, {\bf p}-, and {\bf b}-axes of earthquake focal
mechanisms. Each quadrant of the focal sphere is subdivided into 110
spherical triangles and quadrilaterals with equal area (consequently
covering equal solid angles), and we calculate the number of times a
centroid in projected into these cells. We normalize the numbers, so
that the total number of the entries of Table~1 equals 11,000.
Therefore, if the distribution of centroids in the {\bf tpb}-system
were completely random, all the numbers in the table would be equal
to 100. The entries `Col.~Angle' give the colatitude angle
corresponding to the lower edge of a segment consisting of one or
several cells. The number of pairs for each segment is shown in
`Pair number' column.

Table~1 shows a strong concentration of centroids near the plane
going through the {\bf t--p} and the {\bf b}-vectors. Therefore this
plane should correspond to the fault-plane and that going through
the {\bf t+p} and {\bf b}-vectors should usually correspond to the
auxiliary plane. Most of the strong earthquakes are concentrated in
subduction zones. Fig.~3 in Kagan and Jackson (1994) illustrates
that the fault-plane should include both the {\bf t--p} and the {\bf
b}-vectors.

Huc and Main (2003) suggest that the vertical errors in centroid
determination would strongly influence the direction statistics
between the pairs of events. Even for $M6$ earthquakes, about 1/3 of
shallow centroids are assigned the depth of 15~km. This means that
the depth could not be accurately evaluated.

These results and those shown in Table~1 by Kagan (1992b) support
the conventional model of an earthquake fault: a rupture propagating
with slight deviations along a fault-plane. Figs.~\ref{s_cal3b} and
\ref{s_cal4} suggest that an earthquake occurring on such a fault
system should have the CLVD component close to zero even if subevent
focal mechanisms are significantly disoriented by rotation around
the axis normal to the fault-plane ({\bf n}-axis).

\subsection{Rotation angle statistics }
\label{YYK_stat3}

\begin{figure}
\begin{center}
\includegraphics[width=0.60\textwidth,angle=0]{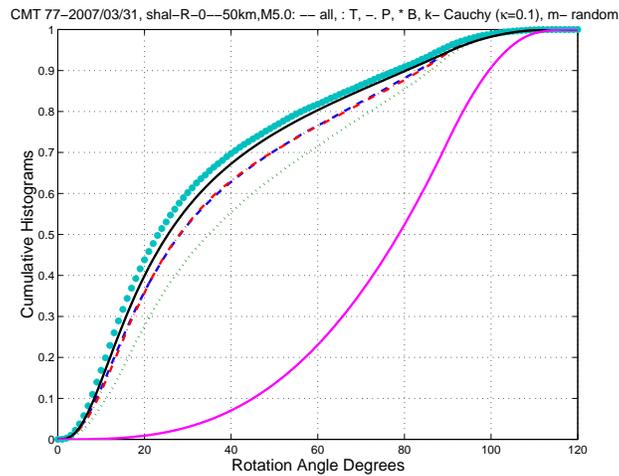}
\caption{\label{s_cal7} Distributions of rotation angles for pairs
of focal mechanisms of shallow earthquakes (depth 0-70~km) in the
CMT catalog 1977/01/01--2007/03/31; centroids are separated by
distances between 0-50~km, magnitude threshold $m_w = 5.0$. Circles
-- all centroids ; crosses -- centroids in 30$^\circ$ cones around
the {\bf t}-axis; plusses -- centroids in 30$^\circ$ cones around
the {\bf p}-axis; stars -- centroids in 30$^\circ$ cones around the
{\bf b}-axis. Left solid line is for the Cauchy rotation with
$\kappa = 0.1$; right solid line is for the random rotation. }
\end{center}
\end{figure}

Fig.~\ref{s_cal7} displays cumulative distributions of the rotation
angle $\Phi$ for shallow earthquake pairs separated by a distance of
less than 50~km. We study whether the rotation of focal mechanisms
depends on where the second earthquake of a pair is situated with
regard to the first event. Thus we measure the rotation angle for
centroids located in 30$^\circ$ cones around each principal axis
(see curves, marked the {\bf t}-, {\bf p}-, and {\bf b}-axes) of the
first event.

The curves in Fig.~\ref{s_cal7} are narrowly clustered, and are
obviously well approximated by the DC rotational Cauchy distribution
(Kagan, 1990, 1992b). This distribution is characterized by a
parameter $\kappa$; a smaller $\kappa$-value corresponds to the
rotation angle $\Phi$ concentrated closer to zero. Thus, all
earthquakes, regardless of their spatial orientation, have focal
mechanisms similar to a close event. Earthquakes in the cone around
the {\bf b}-axis correspond to a smaller $\kappa$-value than the
events near the other axes. These results are similar to the results
shown in Fig.~6 by Kagan (1992b).

\begin{figure}
\begin{center}
\includegraphics[width=0.60\textwidth,angle=0]{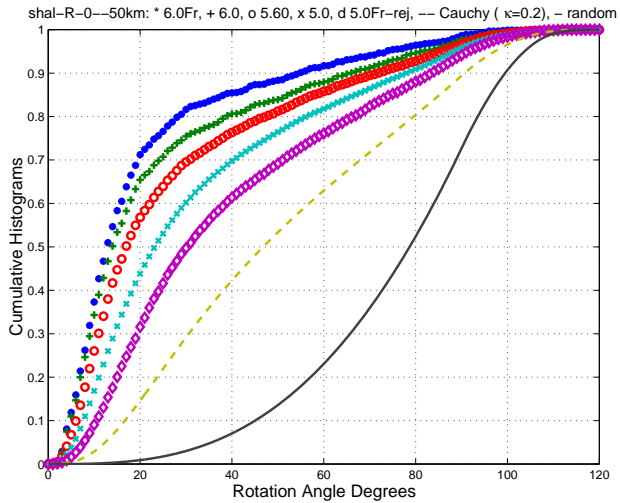}
\caption{\label{s_cal8} Distributions of rotation angles for pairs
of focal mechanisms of shallow earthquakes in the CMT catalog;
centroids are separated by distances between 0-50~km; centroids in
30$^\circ$ cones around the {\bf b}-axis. Stars -- CMT
well-constrained earthquakes, magnitude threshold $m_t = 6.0$,
1977/01/01--2005/01/01; plusses -- CMT catalog, magnitude threshold
$m_t = 6.0$, 1977/01/01--2007/03/31; circles -- CMT catalog,
magnitude threshold $m_t = 5.6$, 1977/01/01--2007/03/31; crosses --
CMT catalog, magnitude threshold $m_t = 5.0$,
1977/01/01--2007/03/31; diamonds -- CMT not well-constrained
earthquakes, magnitude threshold $m_t = 5.0$,
1977/01/01--2005/01/01. Left dashed line is for the Cauchy rotation
with $\kappa = 0.2$; right solid line is for the random rotation. }
\end{center}
\end{figure}

In Fig.~\ref{s_cal8} we show how the distribution of the rotation
angle $\Phi$ depends on the magnitude threshold and the selection of
the well-constrained earthquakes in the catalog
(Section~\ref{YYK_stat1}). It is clear from the diagram that 1) for
larger earthquakes the angle distribution is generally more
concentrated near zero, and 2) the well-constrained earthquakes also
have a smaller angle between pairs. These conclusions are easily
explained by the higher accuracy of strong event solutions (Kagan,
2003): the disorientation of the DC solutions is caused mostly by
solution errors. Kagan (2000) indicates that the rotation angle
error is on the order $10^\circ$ for the best solutions.
Fig.~\ref{s_cal4} suggests that if subevent constituents of an
earthquake have small rotation ($\Phi \le 10^\circ$), the CLVD
component would be close to zero.

Below we show the pattern analysis for the most accurate solutions:
the well-constrained earthquakes $m \ge 6.0$. The disadvantage here
is their small number.

\subsection{Rotation pole statistics }
\label{YYK_stat4}

\begin{figure}
\begin{center}
\includegraphics[width=0.60\textwidth,angle=0]{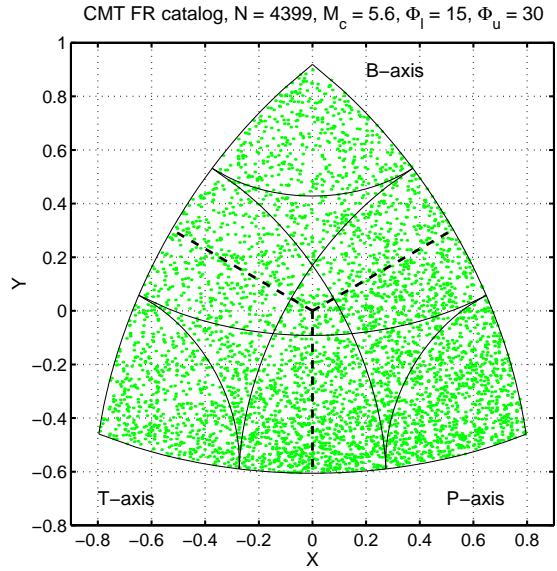}
\caption{\label{s_cal11} Distributions of rotation poles for pairs
of focal mechanisms of shallow well-constrained earthquakes in the
CMT 1977/01/01--2005/01/01 catalog. Centroids are separated by
distances between 0-50~km; magnitude threshold $m_t = 5.6$; the
rotation angle $15^\circ \le \Phi \le 30^\circ $. }
\end{center}
\end{figure}

From Figs.~\ref{s_cal4}--\ref{s_cal5} one can infer that the
rotation pole position strongly influences the CLVD component of a
complex source. In Fig.~\ref{s_cal11} we show the distribution of
the rotation poles for the second earthquake focal mechanism on a
reference sphere of the first event. Because of the symmetry of the
DC source, we reflect the point pattern at our reference sphere at
the planes perpendicular to all axes. Thus, the distribution can be
shown on an octant of a sphere. We use the Lambert azimuthal
equal-area projection (\ref{Eq_g8a}).

\begin{figure}
\begin{center}
\includegraphics[width=0.60\textwidth,angle=0]{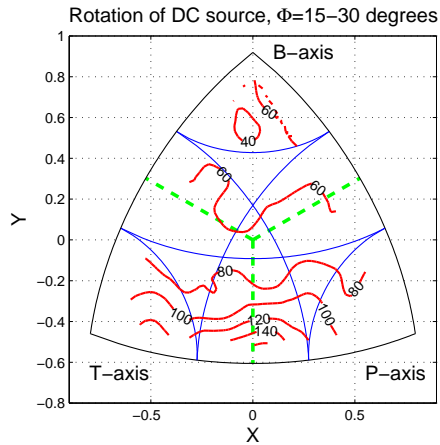}
\caption{\label{s_cal12} Map of rotation poles for pairs of focal
mechanisms of shallow well-constrained earthquakes in the CMT
catalog, see Fig.~\ref{s_cal11}. }
\end{center}
\end{figure}

For example, if in Fig.~\ref{s_cal11}, a pole is shown near the {\bf
b}-axis, this would mean that in both mechanisms the axis has almost
the same orientation: the second mechanism is rotated by an angle
$\Phi$ around an axis intersecting the sphere at the pole. The same
pattern occurs for the poles that are close to other axes. In the
diagram, the angle is $15^\circ \le \Phi \le 30^\circ $. The points
seem to concentrate between {\bf t}- and {\bf p}-axes.
Fig.~\ref{s_cal12} shows the point density distribution, again
confirming that the greatest pole concentration is located near a
nodal-plane.

In Table~2 we display the distributions of the rotation poles on a
reference sphere of the first event. In this Table, as in Table~1,
we subdivide the positive octant of the sphere into 55 spherical
triangles and quadrilaterals with equal area and then calculate the
number of times the rotation axis intersects each of these cells.

The upper chart in Table~2 shows the distribution of axes for the
rotation angles of less than 15$^\circ$. The distribution is
randomly uniform as can be expected, because these rotations are
caused, most probably, by random errors. For larger angles (the
second chart, $15^\circ \le \Phi \le 30^\circ $) the distribution
shows that the stresses of the first earthquake influence the focal
mechanism of the second earthquake in a pair.

To illustrate the data and our technique in Table~3, we show the
calculations for 18 pairs of earthquakes in the lower-left corner
cell of the second chart in Table~2. Earthquakes are distributed
more or less uniformly at subduction zones. We can see again that
many of them have been assigned the standard depth of 15~km. As
explained above in this Subsection, the {\bf t}-axes of both
solutions are almost identical. Therefore, the rotation poles are
concentrated near the axis projection, both in Table~2 and in
Fig.~\ref{s_cal11}. Because the rotation angle is relatively small
($15^\circ \le \Phi \le 30^\circ $) the {\bf p}-axes also have close
orientation.

However, the third and fourth charts of the Table calculated for the
large angles of rotation display a different behaviour. The rotation
poles are concentrated near the {\bf b}- and {\bf p}-axes.
Inspection of these earthquake pairs indicates that 20 of 41 duos
are concentrated in an area 200 by 200~km near the New Hebrides
Islands -- a region of complex subduction tectonics. These
earthquakes likely result from the complicated slab geometry.

We note that in Table~7 in Kagan (1992b) which is similar to
Table~2, the plane orthogonal to the {\bf b}-axis was incorrectly
rotated by 90$^\circ$ and the results collapsed on an octant.
Therefore, the number pattern at the former Table rows at {\bf p}-
and {\bf t}-axis is a mixture of both distributions.

\section{Discussion }
\label{YYK_disc}

We studied the distribution of the non-DC (CLVD) component for a
composite earthquake source or a fault system. A source is
considered complex if it consists of two or more events with a DC
focal mechanism differently oriented. The theoretical computations
detail conditions under which such a source would produce a non-zero
CLVD component. For most of the geometric barriers proposed as
common features in an earthquake fault system (King, 1983, 1986),
the CLVD component should be zero. Frohlich {\it et al.}\ (1989) and
Frohlich (1990) came to similar conclusions.

We tried to statistically estimate whether the geometrical pattern
leading to a zero CLVD could be confirmed by analyzing an earthquake
catalog of focal mechanism solutions. However, the results shown in
the previous Section are not fully convincing. They do, however,
suggest limited support for this conjecture.

Why can't analysis of earthquake focal mechanisms in available
catalogs definitively explain the CLVD component presence? To
understand shear rupture properties, we need to investigate the
behavior of the seismic moment tensor sums for distances between
events close to zero. At such distances, the intrinsic geometrical
conditions of the rupture would play a major role. At larger
distances the slab geometry and its interaction with the upper
mantle would significantly influence the focal mechanisms and
pattern of earthquake hypocenters/centroids. Moreover, because of
the low frequency waves used in the seismic moment inversion,
centroids in the CMT catalog have low accuracy both in the
horizontal plane and at depth. From Smith and Ekstr\"om's (1997) and
Kagan's (2003) results, horizontal accuracy in a centroid location
can be estimated on the order of 15-20~km. Centroid depth
uncertainty should be the same or higher. In our computations we use
distances between the events, which should increase the random error
but decrease systematic uncertainties. The accuracy values quoted
above are comparable with the slab thickness. This means that our
distributions are influenced strongly by slab geometry.

Brudzinski {\it et al.}\ (2007) studied slab geometry for the
subduction zones using higher accuracy (less than 10~km, see above)
hypocentral global earthquake catalogs. They found that earthquakes
concentrate at two layers: double Benioff zones or the upper and
lower boundaries of a subducting slab. This means that large
earthquakes which are listed in the CMT catalog are mostly connected
with the subducting plate geometry, not with any local geometry of
the rupture zone. The geometry of deforming thin elastic sheets of
material is a complex problem even if the sheet is deforming in free
space (Marder {\it et al.}, 2007).

For earthquakes occurring at the double Benioff zones, the treatment
of focal mechanisms should be modified because their symmetry
properties differ from those of earthquakes in the middle of a slab.
For DC sources of earthquakes at the slab boundary, we should be
able, in principle, to identify not only which of the focal planes
is a fault-plane, but also to resolve the ``up and down" of the
fault-plane (Kagan, 1990, p.~576). Hence, the only symmetry
operation for such a focal mechanism is the identity (\ref{eq3a1}).
In such a case, the quaternion visualization techniques described by
Hanson (2005, Ch. 21-23) can be implemented to picture both a slab
surface geometry and the frames of earthquake rupture associated
with the surface. However, the techniques proposed above are
developed for smooth rotations. As we indicated earlier
(Section~\ref{intr2}) earthquakes occur on fractal sets and the
orientation of their focal mechanisms is also controlled by a
non-smooth fractal distribution. Therefore, new methods need to be
created for displaying earthquake focal mechanisms and their spatial
pattern.

While investigating earthquake spatial distribution (Kagan, 2007a),
we were able in some degree to make a smooth transition from a 3-D
distribution of hypocenters for small distances to a 2-D epicenter
distribution for distances comparable or larger than the thickness
of seismogenic crust. After appropriate adjustments, the obtained
values of fractal dimension are comparable for both cases. It seems,
however, that the methods employed for simple point distributions in
2-D and 3-D are much more difficult to implement for tensor
quantities. The challenges are two-fold: theoretical problems of
displaying and interpreting the 6-D distributions and the more
practical problem of no high quality, massive datasets of focal
mechanism solutions.

Several extensive local catalogs of focal mechanism solutions have
been compiled (Pasyanos {\it et al.}, 1996; Kubo {\it et al.}, 2002;
Hardebeck, 2006; Clinton {\it et al.}, 2006; Matsumoto {\it et al.},
2006). They contain thousands of events. Many of these occur near
local seismographic stations and, therefore, their solutions have
good depth control. However, the largest catalogs are based on
first-motion analysis and their solutions have a significantly lower
accuracy than do catalogs of the moment-tensor inversions (Kagan,
2002, 2003). The latter catalogs are not yet sufficiently extensive,
well-documented, or tested to be used for rigorous statistical
studies. However, it is possible that in a few years these catalogs
will be improved.

\section*{Acknowledgments}
I appreciate partial support from the National Science Foundation
through grants EAR~04-09890 and EAR-0711515, as well as from the
Southern California Earthquake Center (SCEC). SCEC is funded by NSF
Cooperative Agreement EAR-0106924 and USGS Cooperative Agreement
02HQAG0008. I thank Kathleen Jackson for significant improvements in
the text. Publication 0000, SCEC.

% \end{article}

\clearpage

\newpage

\newpage

\newpage

\newpage

\newpage

\newpage

\newpage

\newpage

% DISK$USER:[KAGAN.D2C]
%
% FOCM_TABLE1.TXT;5      4/9       25-NOV-2007 17:41:33.99  P46, p.37a
% FOCM_TABLE2.TXT;4     10/18      14-DEC-2007 17:48:09.66  P46, p.38 well-constrained
% % FOCM_TABLE2A.TXT;6    10/18      14-DEC-2007 18:15:37.63  P45, p.49; P46, p.17
% FOCM_TABLE3.TXT;6      6/9       12-DEC-2007 18:11:08.26  P46, p.41

%\end{article}   %disable in DRAFT

\end{document}